\documentstyle[twocolumn,aps]{revtex}

\begin{document}
\baselineskip=12pt
\draft

\preprint{}

\title{Hall Coefficient in the doped Mott Insulator}

\author{Pinaki Majumdar and H. R. Krishnamurthy$^{\dagger}$}
\address{ Department of Physics, Indian Institute of Science, Bangalore
560 012, India. \\
$^{\dagger}$ Also at Jawaharlal Nehru Center
for Advanced Scientific Research,
Jakkur,\\
Bangalore 560 064, India.}

\date{\today}

\maketitle

\begin{abstract}
We compute the (zero frequency)
Hall coefficient $R_H$, and the high frequency Hall
constant $R_H^*$ for the strong coupling Hubbard model away
from half-filling, in the $d=\infty$/ local approximation,
using the new iterated perturbation scheme
proposed by Kajueter and Kotliar. We find quantitative
agreement of our $R_H^*$ with the  QMC results
obtained in two dimensions by Assaad and
Imada [Phys. Rev. Lett. {\bf 74}, 3868 (1995)]. However
while the {\it sign} of $R_H$ is
quite accurately reproduced by $R_H^*$ the doping dependence
of its magnitude at large U is not.
We report results for
the complete temperature $(T)$, doping $(x)$ and
$U$ dependence of $R_H$ and $R_H^*$ and discuss
their possible relevance to doped cuprates.
\end{abstract}

%\newpage

\pacs{PACS numbers: 71.28.+d,71.30.+h,72.10-d}

\narrowtext

%lesssim,gtrsim

%\maketitle

Hall measurements are normally regarded as
a standard probe for the density
and {\it sign} of charge carriers in electronic systems.
Within a Boltzmann, continuum $(\epsilon_{\vec k}
\sim k^2)$ picture, the Hall coefficient, $R_H$, is
given by $1/ne$, where $n$ is
the density and $e$ the charge of the carriers. In the presence
of a lattice, and for weak scattering,
$R_H$ depends on the curvature
($\partial^2 \epsilon_{\vec k}/\partial k^2_x$, etc)
of the Fermi surface (FS)~\cite{ziman}
and need not have the same sign
as $e$; an "electron-like" (positive
curvature) FS
implies
$R_H < 0$
and a "hole-like" (negative curvature)
FS implies $R_H>0$.

An obvious question of great interest is  whether this
correspondence would survive for  strong correlations.
Experiments
on some of the high $T_c$ systems \cite{ong1,batlogg}
reveal a hole-like FS~\cite{photoem}
while $R_H$ changes from electron-like to
hole-like with electron doping~\cite{uchida}. In addition, both
the doping and temperature dependence of $R_H$
are "anomalous":
$R_H$ shows unbounded growth as the "half-filled" state
is approached, while the temperature
dependence is roughly $1/T$, with a saturation
as $T\rightarrow T_c$, the superconducting transition
temperature, from above. The physics of the cuprates is believed
to be governed by  strong electron-electron repulsion (large $U$,
in the Hubbard context) and low dimensionality $(d=2)$.
This has motivated the search for   controlled
approximations that help to
isolate the generic correlation
affected features in transport in strongly correlated models.
For the Hubbard model in $d=2$ the
only controlled calculations that exist are for the
high frequency
Hall constant, $R_H^*$, defined by
\cite{sss,asad,mahesh}
\begin{equation}
R_H^* = \lim_{B\rightarrow 0}
\lim_{\omega \rightarrow \infty}
\sigma_{x,y}(\omega)/[B\sigma_{x,x}^2(\omega)]
\end{equation}

Unlike the usually measured (zero frequency) Hall
coefficient, $R_H$, which involves
low energy scattering processes, $R_H^*$ defined above
involves only the reactive response. Nevertheless, as demonstrated
first by Shastry {\it et al} \cite{sss}, $R_H^*$ encodes
information about correlation effects in the system and
(unlike $
R_H$) is accessible within controlled calculations.
For the Hubbard model it has been computed
through  Quantum Monte Carlo \cite{asad} and
finite cluster \cite{mahesh} calculations,
and for the $t-J$ model through high temperature
series expansion
\cite{sss}, and cluster
calculations \cite{mahesh}. Finite size effects
and analytic continuation problems
which hinder low frequency transport calculations
in these contexts do not significantly
affect the results on $R_H^*$.

One significant result
of these studies is the finite temperature crossover from
electron-like to hole-like  $R_H^*$ as the hole doping
concentration, $x$,
away from half-filling
is reduced below $\sim 30\%$. This is consistent
with some of the high $T_c$ data on $R_H$~\cite{ong1,batlogg}
and raises the intriguing possiblity
that  $R_H^*$ may mimic all the features of
$R_H$.
Hence
it would be interesting
to calculate {\it both} $R_H$ and $R_H^*$
for a strongly correlated model within
a single scheme and clarify the connection between them.
The recently developed
$d=\infty$
or local approximation ~\cite{metzvoll,gkprb}
does provide us  with
a controlled scheme for studying
this connection in a strongly
interacting system.
We report and discuss the results of such a study
of the $R_H$ and  $R^*_H$
for the doped Mott insulator phase of the
Hubbard model in this paper. Our primary results
can be summarised as follows.$(i)$ The
results for  $R_H^*$ computed in the local
approximation are quantitatively consistent
with QMC results~\cite{asad} obtained for d=2. $(ii)$ There is
a {\it complete correspondence} of the {\it sign} of $R_H$
and $R_H^*$
computed within this approximation at {\it intermediate} $T$
over a wide range of $x$ and $U$.
$(iii)$ However, $R_H$ and $R^*_H$,
differ in sign as $ T\rightarrow 0$,
and their {\it magnitudes} have {\it qualitatively different}
dependences on $x$, for intermediate
$T$ and large $U$, close to half-filling.

Our calculations are based on the
(by now well known) mapping of Hubbard model in the
$d\rightarrow\infty$
limit to a self-consistently
embedded impurity problem~\cite{metzvoll,gkprb}.
At half-filling this model shows~\cite{motttrans}
a transition from
a Fermi-liquid to a Mott insulating phase at $U=U_c \sim 2.5$
(in units of D, the half-bandwidth, which we set to 1) at $T=0$,
as long as there  is sufficient frustration in the lattice to
suppress magnetic order.
Away from half-filling the system
is metallic, and there is no evidence of magnetic order
\cite {jarfree}. Most of this has been learnt~\cite{rmp} using a
combination of Quantum Monte Carlo, exact diagonalisation,
the NCA~\cite{hewson}, and iterated perturbation
theory (IPT)~\cite{gkprb,gkrauthprb,rozprb}
schemes for solving the
impurity problem. Of these, the IPT scheme, because of its
simplicity, has been particularly useful
in clarifying the physics at half-filling.
It reproduces Fermi-liquid behaviour for $U<U_c$
and is exact in both the atomic $(t=0)$ and the band $(U=0)$
limits.
Although it does not reproduce the exact critical behaviour
as $U\rightarrow U_c$ at $T=0$, it gives excellent results at
intermediate temperatures and
allows one to directly calculate the
continuous {\it real-frequency}
spectra of the system.
The absence of a similar approximation away from
half-filling,
where the proximity to
a correlation driven insulating state is expected to
profoundly affect the "metallic" behaviour, had
until recently been a
bottleneck.
A recent paper by Kajueter and Kotliar
\cite {modipt} achieves a significant advance
by devising an interpolation
scheme which is exact in the atomic and
band limits at any filling and obtains the correct low energy
physics by imposing the Friedel sum rule.
Our calculations are based on this new scheme,
which as originally proposed was for $T=0$, but
has been adapted by us to $T\neq0$ in a straightforward
way.

We refer the reader to
the original paper ~\cite{modipt} for the detailed formulation
of this scheme and comparison with exact results. For
our purposes it is enough to note that in this scheme
the self-energy  at
$T=0$ is approximated as
$\Sigma_{int}(\omega)=
A\Sigma_2(\omega)/[1-B\Sigma_2(\omega)]$
where $\Sigma_2$ is the analytically continued
second order self-energy
\begin{equation}
\Sigma_2(\omega)=
\left[ {U^2\over{\beta}^2} \sum_{n_1,n_2}
{\cal G}_{n - n_1}
{\cal G}_{n_2}
{\cal G}_{n_1+n_2} \right]_
{i\omega_n \rightarrow \omega^+}
\end{equation}
with ${\cal G}_n\equiv {\cal G}(i\omega_n)$  etc.
Here $A\equiv{n(1-n)\over n_0(1-n_0)}$ and
$B\equiv{(1-n)U -\mu + \tilde\mu_0 \over
n_0(1-n_0)U^2}$
have been chosen so as to reproduce the atomic limit and the
high frequency behaviour of the self-energy at
any filling;
$n\equiv\int_{-\infty}^0 \rho_G(\omega) d\omega$ and
$n_0\equiv\int_{-\infty}^0 \rho_{\cal G}(\omega) d\omega$,
where $ \rho_G(\omega) $ and
$\rho_{\cal G}(\omega)$ are the spectral functions of
the full local propagator $G(\omega)$
and the "bare" local propagator ${\cal G}(\omega)$ respectively;
${\cal G}^{-1} (\omega)=\omega + \tilde\mu_0 -\Delta(\omega)$,
where $\Delta(\omega)$ is the hybridisation
of the self-consistently embedded impurity problem and
$\tilde\mu_0$ is a free parameter  fixed by
imposing the
Friedel sum rule.
At $T=0$ and
for a given $U$, the chemical potential $\mu$, "quasi chemical
-potential" $\tilde\mu_0$ and occupation number $n$ are related
uniquely through the interpolation and sum rule constraints.

The ground state averages $n$ and
$n_0$  which enter the interpolating
self-energy $\Sigma_{int}(\omega)$ trivially generalise to
$T\neq 0$ as thermal averages
$(n = \langle\hat{n}\rangle
=\int_{-\infty}^\infty \rho_G(\omega)f(\omega) d\omega$,
$f\equiv[exp(\omega/T) + 1]^{-1};$,
etc )
and still yield a sensible
interpolation.
For $T\neq 0$ and a desired $n$
we fix $\tilde\mu_0(n,T)$ as $\tilde\mu_0(n,T=0)$ and
solve for $\mu(n,T)$
self-consistently.
Thus once the connection between
$\mu,n$ and $\tilde\mu_0$
is established at $T=0$
for a given $U$, the $T\neq 0$ calculations only
involve the $d=\infty$ consistency loop and are computationally
quite affordable. We compute the  self-energy, $\Sigma(\omega)$
of the
lattice problem as the self-consistent self-energy
$\Sigma_{int}(\omega)$ of
the impurity problem directly for real frequencies,
and use it to study $R_H$ and $R^*_H$
for large U, ($U/D\sim 4-6$)
as a function of $T$ and $x$.
$R_H$ is computed from the
conductivity tensor $\sigma_{\alpha\beta}(\omega=0)$.
In the $d\rightarrow\infty$/local approximation the transport
coefficients do not involve vertex corrections~\cite{khurana}
and the dc conductivity  is given by
\begin{equation}
\sigma_{xx}(0) =c_{xx} \int d\epsilon \rho_0 (\epsilon)
\int d\omega
A^2 (\epsilon,\omega) ({-\partial f/\partial\omega})
\end{equation}
where
$c_{xx} = e^2\pi/(d \hbar a_0)$,$a_0$ is the lattice
spacing,
$\rho_{0}(\epsilon)=(2/{\pi D})\sqrt{1 -
(\epsilon/D)^2}$ is
the bare density of states (DOS),
and
$A(\epsilon,\omega)\equiv -{1\over\pi} Im[\omega + \mu -\epsilon
-\Sigma(\omega)]^{-1}$ is the spectral function.
The transverse conductivity is given by~\cite{jarrev,hallform}
\begin{equation}
\sigma_{xy}(0) = B c_{xy}\int d\epsilon  \rho_0 (\epsilon) \epsilon
\int d\omega
A^3 (\epsilon,\omega) ({-\partial f/\partial\omega})
\end{equation}
where $c_{xy} = {\vert e\vert}^3 \pi^2 a_0/(3d^2 \hbar^2)$.
$R_H$ is then given by
$\sigma_{xy}(0)/[B\sigma_{xx}^2(0)]$.
For $R^*_H$ we use the expression for a
hypercubic lattice  \cite  {sss}
\begin{equation}
R^*_H=({a_0^3}/\vert e\vert)
{\sum cosk_xcosk_y\langle n_{k,\sigma}
\rangle/[\sum cosk_x\langle n_{k,\sigma}
\rangle]^2 }
\end{equation}
In the spirit of the local
approximation we compute $\langle n_{k,\sigma}
\rangle$, the momentum distribution function,
from our self-energy and do the  ${\vec k}$ sums over
a 2d Brillouin zone.

We first discuss $R^*_H(x,T)$  since  2d QMC
\cite {asad} and cluster calculations~\cite{mahesh} provide us
with results to compare with.
For large $U$, $(U/t \sim 16)$
QMC results are
available only above $T/t\sim 0.25$, ($T\sim 0.06$ in our
units, $D\sim 4t$)
and there is a clear peak structure in $R^*_H(T)$
with the location of the
maxima $T^{QMC}_{max}$ gradually shifting to lower $T$ as
$U$ increases at a fixed doping $(x=0.05)$. This is
consistent with our results; but as $T\rightarrow 0$ we
find an "anomalous" region $(R^*_H>0)$
in the $U-x$ plane which could
not be accessed within QMC but is clearly seen in cluster
calculations \cite{mahesh}. The detailed comparison is
shown in Fig.1.

Exploring the detailed $x,T$ dependence of $R^*_H$ and
$R_H$ for
$U/D = 4$ (roughly
$U=16t$ in 2d), we find that
$R^*_H$ remains hole-like down to $T=0$
for $x\lesssim 0.12$ (Fig.2a) while for $x\gtrsim 0.20,
R^*_H$ is electronlike for all $T$. $R_H$ however is
{\it electron-like} for all $x$ at $T=0$ (Fig.2b).
This, as we
discuss below, is a consequence of the Friedel sum rule
which forces $R_H$ to
preserve its non-interacting, band structure
determined, value at
$T=0$. For small $x$ however there is a rapid crossover
to a hole-like $R_H$ with increasing $T$.
Fig.3 shows the crossover temperature $T_{cr}$
as located from the numerically evaluated $R_H$.
$T_{cr} \sim x^2$  for small, $x$ (Fig.3), and
rises more steeply as  $x\rightarrow x_c
\sim 0.25$, beyond which there is no crossover.
For $x<x_c\sim 0.25$ there is a high temperature branch
of $T_{cr}(x)$ across which there is a "re-entrant" change
of sign in $R_H$.
For $x\gtrsim 0.25$
$R_H$, like $R^*_H$, is electron-like for all $T$.
While the $T$ dependence illustrates the
similarity in the {\it sign} of $R_H$ and $R^*_H$ at
intermediate $T$ ($T\gtrsim 0.15\sim O(J)$ in 2d) the
$T\rightarrow 0$ result illustrates their essential
difference.
This difference arises from the
fact that, for $T\rightarrow 0$, $R_H$ is a
Fermi surface object, completely dominated by the {\it
coherent
part} of the spectra (which leads to the divergent
$\sigma_{\alpha\beta}$), while $R^*_H$ being a
"high-frequency" object gets
contribution from all over the Brillouin zone.

The fact that at  $T=0$ all correlation effects
cancel out of $R_H$ \cite{kotmott} is easily seen
by using the $\omega \rightarrow~0, T\rightarrow~0$
Fermi liquid parametrisation of the spectral function~:
$A(\epsilon,\omega) \sim
\Gamma/((m^*\omega +\bar\mu -\epsilon)^2 + \pi^2 \Gamma^2)$
where $\Gamma\equiv -(1/\pi) Im \Sigma(\omega=0)$,
$m^*\equiv 1-(\partial \Sigma_R/\partial \omega)_{\omega =0}$
and $\bar\mu\equiv \mu(x,T) - \Sigma_R(x,T, \omega=0)$.
Since $ {-\partial f/\partial\omega}\rightarrow
\delta(\omega)$ as $T\rightarrow 0$,
$\sigma_{xx}\sim \rho_0(\bar\mu)/\Gamma$,
$\sigma_{xy}\sim B \bar\mu \rho_0(\bar\mu)/\Gamma^2$
and $R_H\sim \bar\mu$ so long as
$\Gamma << D$. Since the sum rule
requires $\bar\mu(x,T=0)= \mu(x,T=0,U=0)=\mu_0 (x)$,
$R_H$ is pinned to its band value at $T=0$.
The small $x$ behaviour of the lower $T_{cr}$
(Fig.3) can also be
understood within this phenomenology. For $x\rightarrow 0$
(accessible within our numerics so far
upto $ x\sim 0.06$) $\bar\mu(x,T)$
increases rapidly with $T$. We find $\bar\mu(x,T) \sim
\mu_0 (x) + \alpha T^2/x^3$, where $\alpha>0$ depends on
$U$. Since the free band chemical potential
${\mu_0}(x)\sim -x$
for $x\rightarrow 0$, the crossover scale $T_{cr}\sim x^2$.
The rapid rise of $\bar\mu$ with increasing $T$ is
primarily due to the reduction of $\Sigma_R(x,T,
\omega=0)$.
This in turn is due to the
redistribution of spectral weight
in $Im\Sigma(\omega)$
which rises rapidly at  low frequencies
as $T$ increases, causing the disappearance of the resonant feature
in
$\rho_G(\omega)$  at the Fermi level.

The difference between $R_H$ and $R^*_H$ is further
highlighted in their doping dependence as  $x\rightarrow 0$
at intermediate $T$. $R_H\sim g(T)/x$ with $g(T)$
increasing with $T$. This behaviour is seen down to the
lowest accessible $x$ ($\sim O(0.04)$ at these $T$), despite
particle-hole symmetry which should force $R_H(T)=0$ at
$x=0$. This indicates that due to the large finite
$T$ resistivity in the "metal"  for $x\rightarrow 0$
(and the absence of any coherence/sum rules unlike at $T=0$),
$R_H$ "diverges", as if the doped holes could
be treated as effective carriers within a "Boltzmann like"
scenario. $R^*_H$ on the other hand  shows a {\it peak}
as $x$ reduces and then a clear turnaround, with $R^*_H
\sim x$ for $x\rightarrow 0$. Thus as the "Mott" phase is
approached for $T\neq 0$, $R_H$ is "divergent"
(with possibly
an exponentially small turnaround scale because
$\sigma_{\alpha\beta}$
is finite) while $R^*_H$ is benign. This
feature of $R^*_H$ has been seen within QMC~\cite{asad}
and confirms that  for a one band model  with
particle-hole symmetry $R^*_H$ is regular as $x\rightarrow
0$.
The "divergence" in $R_H$ as $x\rightarrow 0$ had been seen
earlier
within slave-boson  gauge theories based on
the $t-J$ model~\cite{gaugeth}, though they
yield only a hole-like $R_H$ ({\it i.e} no sign change even
as
$x\rightarrow 1$, {\it i.e} $n\rightarrow 0$).
Within the $d\rightarrow \infty$
scheme Pruschke {\it et. al}~\cite{jarrev} have computed
$R_H$ using a combination of QMC and the NCA. Our results
are consistent with the QMC data while avoiding the
problems of analytic continuation and access
low $T$, large $U$ regimes much better.

It is  natural to compare our results with the
$R_H$ data on the cuprates~\cite
{ong1,batlogg,uchida} since the cuprates are believed to
be doped "Mott" insulators. Although certain  features
of the data, such as the
sign change with  increasing $x$~\cite{uchida},
"re-entrance"~\cite{ong2}, and the $1/x$ behaviour of
$R_H$~\cite{batlogg,uchida} are recovered
within the local approximation the $T$ dependence of
$R_H$, and the characteristic temperature scales are
not captured. Even taking into
consideration that the $T=0$ metallic state is not
accessible due to superconductivity, experimentally
$R_H$ decreases (roughly as $1/T$) beyond a  $T$ scale
which  is
$O(50K -100K)$. In our calculation $R_H$ decreases with
increasing $T$ only beyond a temperature scale $\sim O(D/10)$.
As demonstrated by its success in understanding
metal-insulator transition in 3d transition oxides~
\cite {dnflatt}
the local approximation quantitatively captures
the physics of strong local  correlations.
To the extent that one believes that
the low energy physics of the cuprates
can be described by the large $U$ Hubbard model,
one must attribute  deviation
of the measured from the computed  $R_H$ to
one or more of the following features which are
left out in our treatment:
$(a)$ bandstructure
effects, including particle-hole asymmetry
and the proximity to the Fermi level
of the van Hove singularity in the quasi 2d DOS.
$(b)$ the ${\vec k}$ dependence of single particle
self-energy,
and $(c)$ vertex corrections to the conductivity,
which do not vanish when $(b)$ is relevant. What
is nice about the local approximation is that it has
allowed a clean separation of the local correlation effects
on $R_H$, which we have discussed in this paper, from the
effects attributable to these other complications, which
we hope to study in future work.

We thank P. N. Mahesh, B. S. Shastry and T. V. Ramakrishnan
for discussions, and Alexander Punnoose and Prabhakar
Pradhan for help with the figures.

FIG. 1 $R^*_H$ in units of $a_0^3/{\vert e \vert}$,
for different $U/D$.
Triangle $U/D=4$;
inverted triangle $U/D=3$;
cross $U/D=2$. Inset: QMC data \cite {asad}, same
legend for $U/{4t}$.

FIG. 2  $R^*_H$ ($a$) and $R_H$ ($b$) in units of
$a_0^3/{\vert e \vert}$, for $U=4$.

FIG. 3 Crossover temperature $T_{cr}(x) $ for
$U=4$. The line is a quadratic fit for $x\rightarrow 0$.
"normal" and "anomalous" refer to $R_H<0$ and
$R_H>0$ respectively.

FIG. 4  $R_H$ ($a$) and $R^*_H$ ($b$) in units of
$a_0^3/{\vert e \vert}$ for $U=4$.

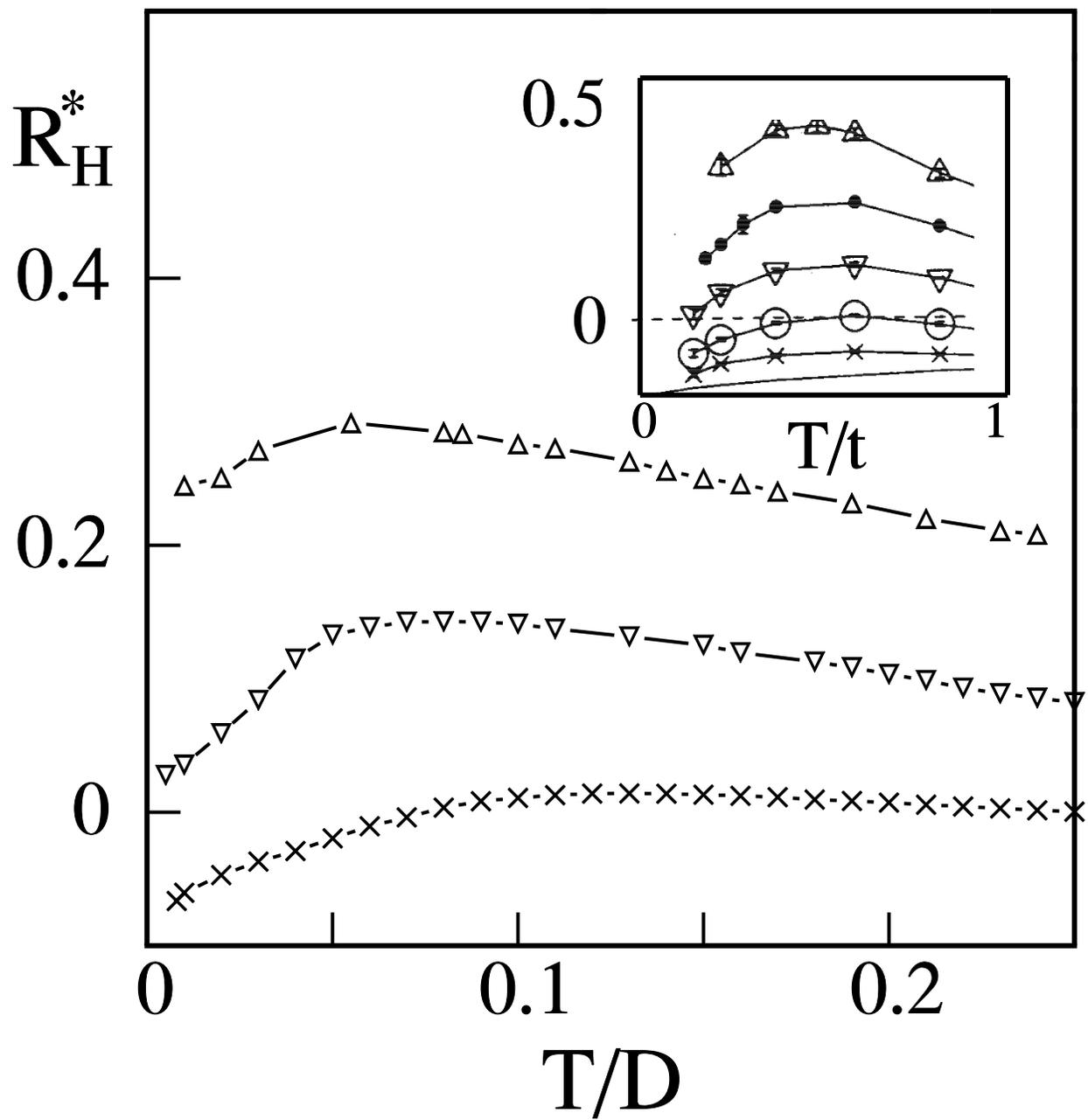
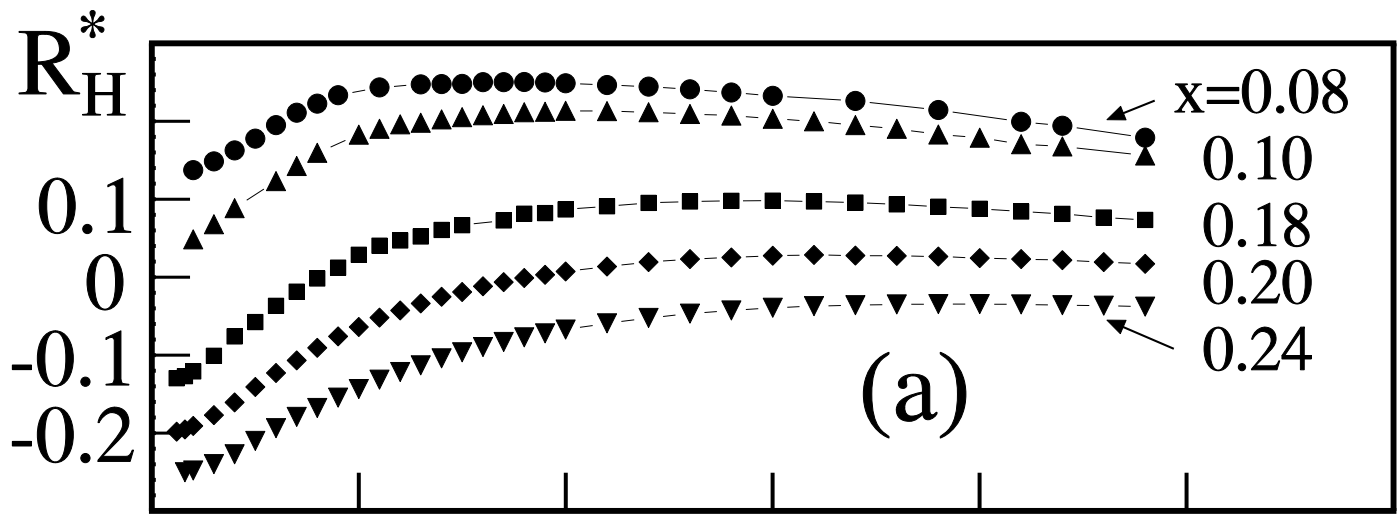
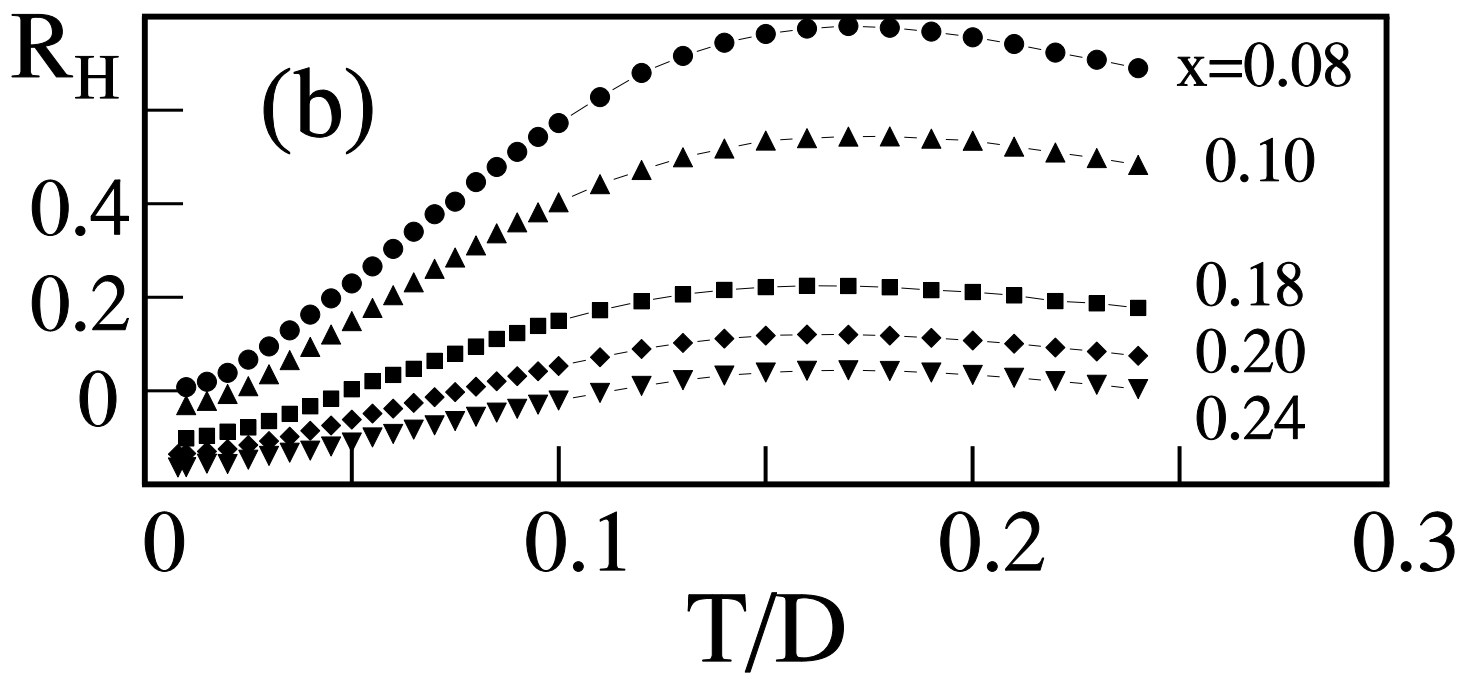
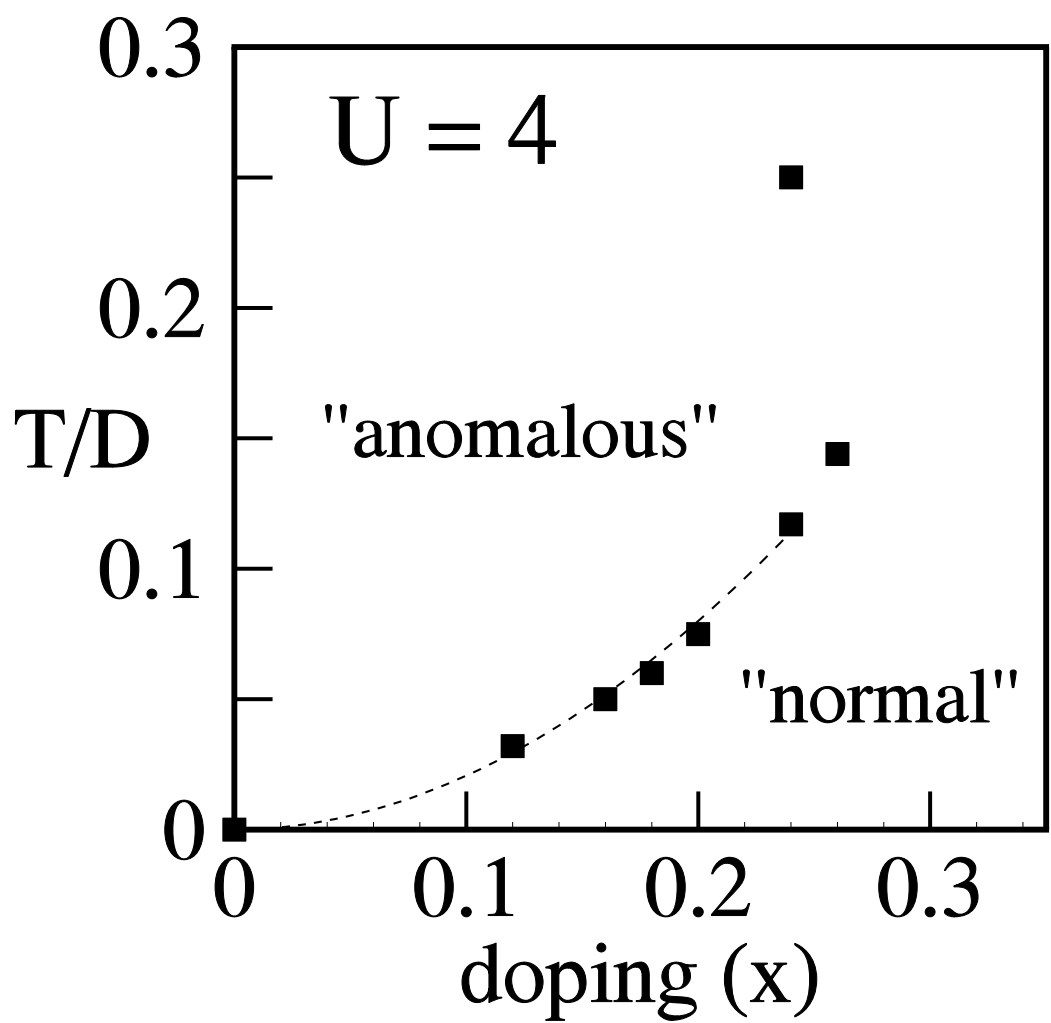
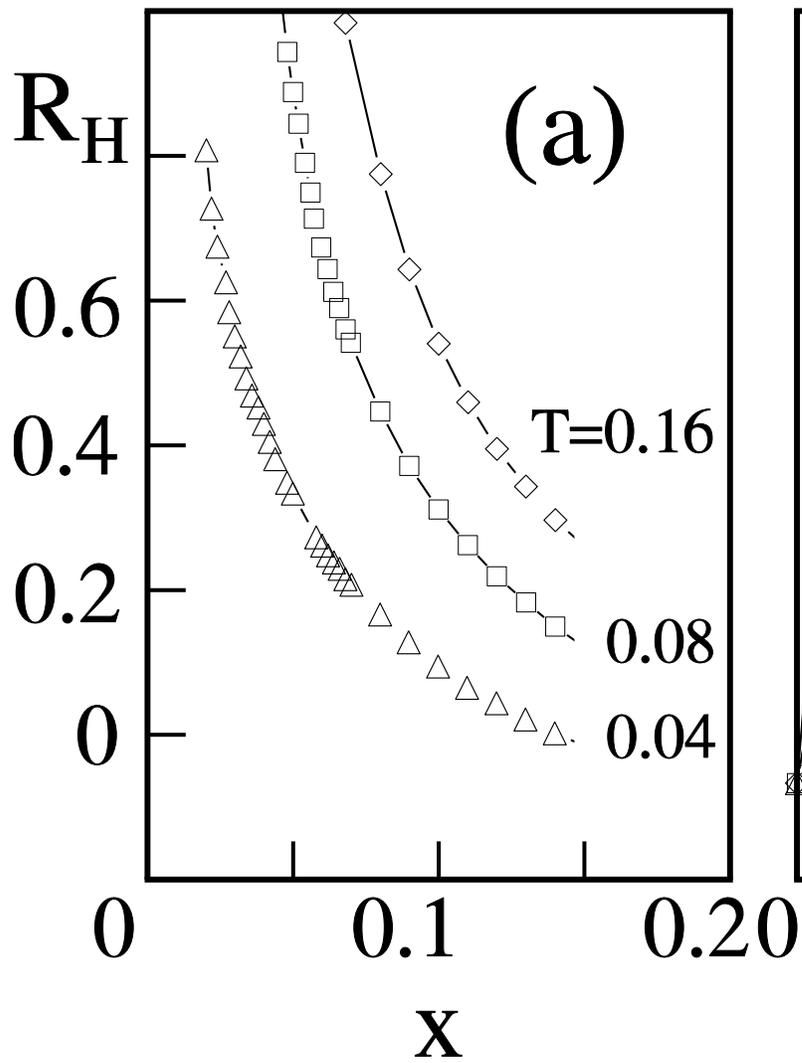
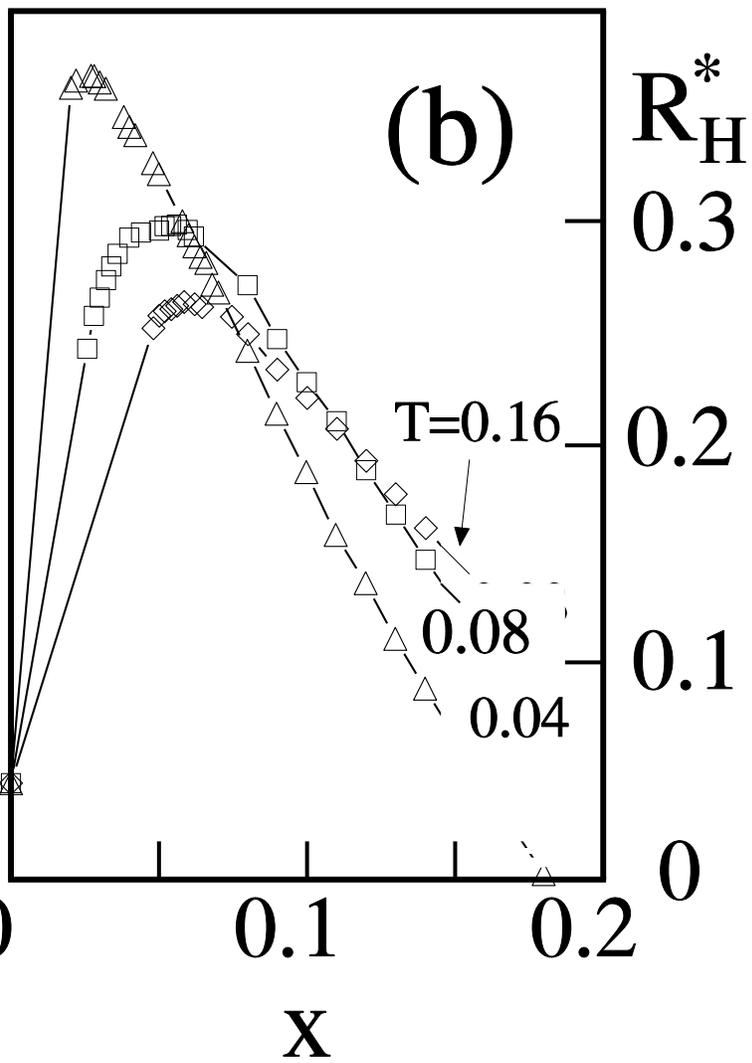
\end{document}